\documentclass[letterpaper, 10pt, conference]{ieeeconf}
\IEEEoverridecommandlockouts	% to override locked commands

\usepackage{multirow}
\usepackage{lipsum}
\usepackage{amsmath}
\usepackage{amssymb}
\usepackage{subcaption}
\usepackage{graphicx}
\graphicspath{{./figures/}}
\usepackage{glossaries}
\usepackage[table,xcdraw]{xcolor}
\usepackage[ruled,vlined]{algorithm2e}
\usepackage[final]{listings}
\usepackage{longtable}
\usepackage{cite}
\usepackage[normalem]{ulem}
\usepackage{booktabs}
\usepackage{comment}
\usepackage{tikz,xcolor,hyperref}
\definecolor{lime}{HTML}{A6CE39}
\DeclareRobustCommand{\orcidicon}{
	\begin{tikzpicture}
	\draw[lime, fill=lime] (0,0) 
	circle [radius=0.16] 
	node[white] {{\fontfamily{qag}\selectfont \tiny ID}};
	\draw[white, fill=white] (-0.0625,0.095) 
	circle [radius=0.007];
	\end{tikzpicture}
	\hspace{-2mm}
}
\foreach \x in {A, ..., Z}{\expandafter\xdef\csname orcid\x\endcsname{\noexpand\href{https://orcid.org/\csname orcidauthor\x\endcsname}
			{\noexpand\orcidicon}}
}

\definecolor{mygray}{rgb}{0.5,0.5,0.4}
\definecolor{mygreen}{rgb}{0.2,0.5,0.2}
\definecolor{myblue}{rgb}{0.2,0.2,0.9}
\definecolor{mykwdclr}{rgb}{0.2,0.6,0.6}

\lstdefinestyle{mybase} {
	language=[Sharp]C,
	breaklines=true,
	showstringspaces=false,
	basicstyle=\small\tt,
	frame=single,
    numbers=left,
    numbersep=8pt,
	numberstyle=\tiny\color{mykwdclr},
	keywordstyle=\color{myblue},
	keywords=[2]{Mathf,Laser_Scanner,MonoBehaviour,Vector3,RaycastHit,Debug,Color,Physics}, 
	keywordstyle=[2]\color{mykwdclr}, 
	commentstyle=\color{mygreen}\ttfamily
}

\lstdefinestyle{mycsh} {
	style=mybase
}

\newacronym{VRU}{VRU}{vulnerable road user}
\newacronym{VRUs}{VRU}{vulnerable road users}
\newacronym{ADAS}{ADAS}{Advanced Driving Assistant Systems}
\newacronym{LKA}{LKA}{Lane Keeping Assistance}
\newacronym{LDW}{LDW}{Lane departure warning}
\newacronym{LCA}{LCA}{Lane Change Assistant}
\newacronym{SAE}{SAE}{Society of Automotive Engineers}

\newacronym{IRI}{IRI}{International Roughness Index}
\newacronym{IMU}{IMU}{Inertial Measurement Unit}
\newacronym{ITS}{ITS}{Intelligent Transportation Systems}
\newacronym{ICC}{ICC}{Intraclass Correlation Coefficient}
\newacronym{GNSS}{GNSS}{Global Navigation Satellite System}
\newacronym{PSR}{PSR}{Present Serviceability Rating}

\title{\LARGE \bf Multi-Modal Assessment of Road Roughness Using Smartphone Applications, Acceleration, and Passenger Ratings}

\author{Novel Certad\orcidN{}\emph{Graduate Student Member, IEEE}, Amirhesam Aghanouri\orcidA{}\emph{Graduate Student Member, IEEE},\\Joseba Gorospe\orcidJ{}\emph{Member, IEEE}, and Cristina Olaverri-Monreal\orcidC{}\emph{Senior Member, IEEE}%
\thanks{Department Intelligent Transport Systems, Johannes Kepler University Linz, Altenberger Straße 69, 4040 Linz, Austria.
\texttt{\{novel.certad\_hernandez, amirhesam.aghanouri, joseba.gorospe, cristina.olaverri-monreal\}@jku.at}}%
}

\begin{document}

\maketitle
\thispagestyle{empty}
\pagestyle{empty}
\captionsetup[figure]{name={Fig.},labelsep=period}
%%%%%%%%%%%%%%%%%%%%%%%%%%%%%%%%%%%%%%%%%%%%%%%%%%%%%%%%%%%%%%%%%%
\begin{abstract}
This paper investigates a multi-modal and human-centric framework for low-cost road roughness assessment. The evaluation was based on three complementary data sources: smartphone-based International Roughness Index (IRI) estimates from two independent smartphone-based applications; in-vehicle GNSS/IMU Receiver (Global Navigation Satellite System Receiver with Inertial Measurement Unit) measurements, and passenger Present Serviceability Ratings (PSR). Data were collected over 1700 km across Austria, Hungary, and Romania under real traffic conditions. Inter-application agreement was evaluated using correlation analysis, Intraclass Correlation Coefficient (ICC), and Bland–Altman methods. While the two smartphone applications show strong correlation, systematic bias limits their interchangeability. A significant inverse relationship between IRI and PSR confirms perceptual sensitivity to roughness, and positive correlations between IRI and vertical acceleration validate the physical linkage between pavement irregularities and vehicle dynamics. The results demonstrate the challenges of integrating consumer-grade sensing and perception-based evaluation for road roughness monitoring as an alternative to high-cost specialized survey equipment.

\end{abstract}

%%%%%%%%%%%%%%%%%%%%%%%%%%%%%%%%%%%%%%%%%%%%%%%%%%%%%%%%%%%%%%%%%%
\section{Introduction}
\label{sec:introduction}

Road surface condition is a fundamental component of transportation system performance, affecting safety, ride comfort, vehicle operating costs, and infrastructure durability. Among functional pavement indicators, the \gls{IRI} is the most widely adopted due to its robustness, standardisation, and transferability across networks and measurement systems \cite{PEREZACEBO2023}. It represents the simulated response of a standardised quarter-car model to the measured longitudinal road profile, providing a consistent measure of ride quality and surface unevenness \cite{XU2024}.

\begin{figure}[t]
	\centering
	\begin{subfigure}{0.237\textwidth}
		\includegraphics[width=\textwidth]{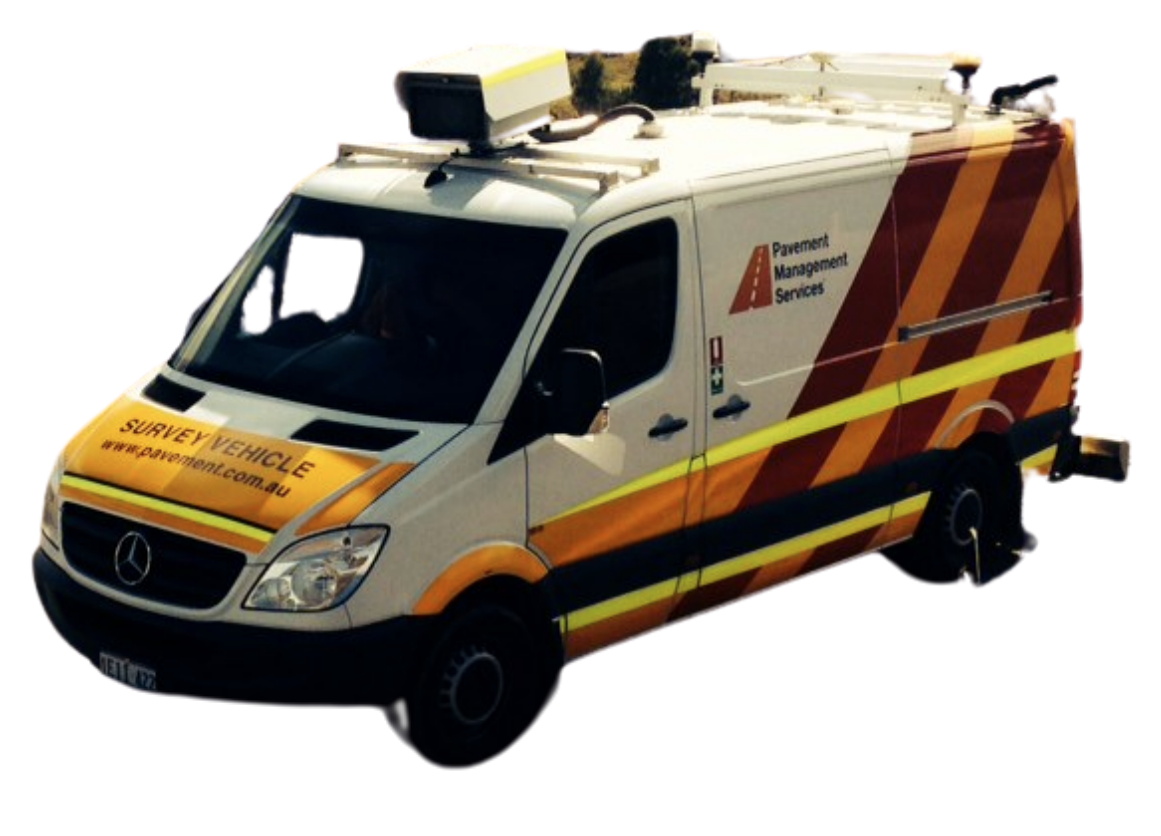}
		\caption{}
	\end{subfigure}
    \begin{subfigure}{0.237\textwidth}
		\includegraphics[width=\textwidth]{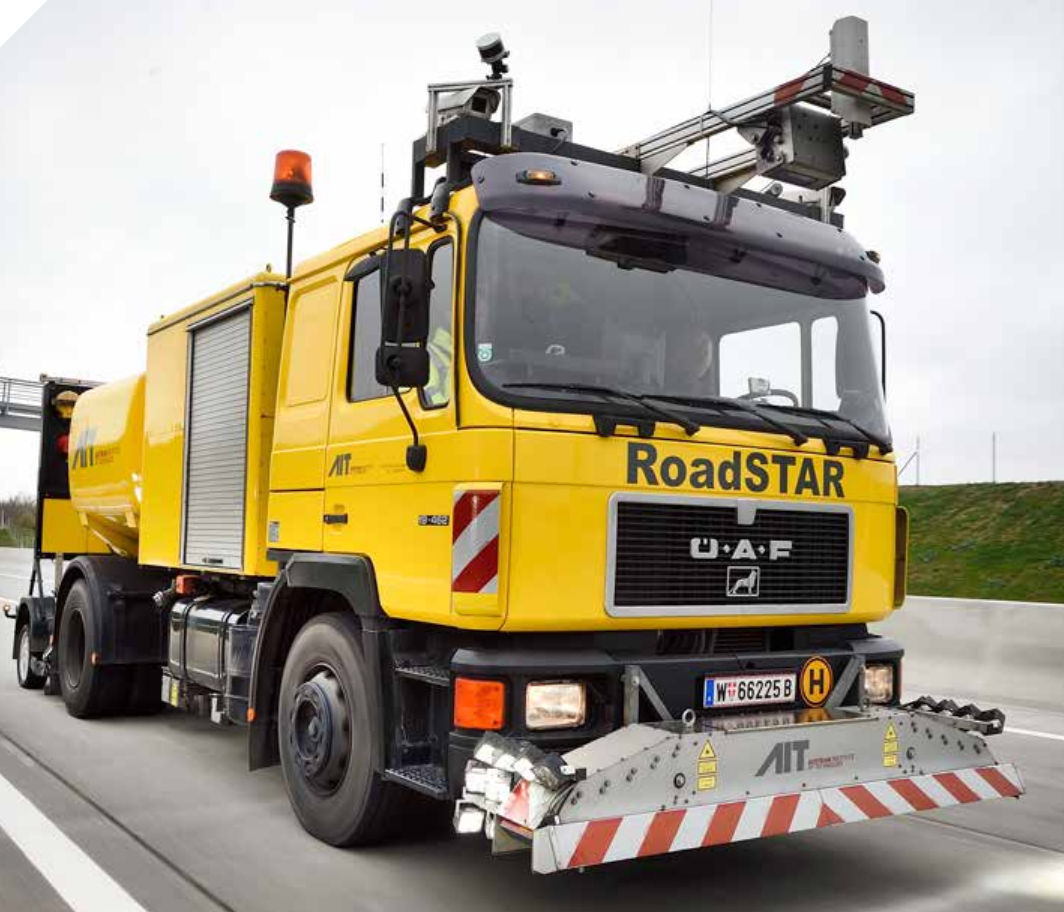}
		\caption{}
	\end{subfigure}
	\caption{Survey vehicles used for high-precision road measurements including \gls{IRI}: (a) ARAN LCMS from Pavement Management Services \cite{PMSARAN}, and (b) RoadSTAR from AIT \cite{AITRoadstar}.}
	\label{fig:survey_vehicles}
    \vspace{-0.5cm}
\end{figure}

\gls{IRI} is traditionally measured using high-speed inertial profilers or automated road analyser vehicles equipped with laser sensors, accelerometers, and distance-measurement instruments (see Fig.~\ref{fig:survey_vehicles}) \cite{XU2024}\cite{PMSARAN}\cite{AITRoadstar}. Although highly reliable, these systems involve substantial acquisition and operational costs \cite{ALATOOM2026}. Consequently, network-level surveys are typically conducted every two to four years, creating a temporal monitoring gap.

This periodic measurement strategy creates a critical gap in pavement monitoring. Pavement conditions may deteriorate due to traffic loads and environmental effects, yet agencies lack updated data to support timely maintenance decisions. This limitation is particularly relevant in the context of \gls{ITS}, where near-real-time infrastructure awareness is increasingly required for data-driven asset management.

To address this challenge, recent research has explored low-cost and portable alternatives for indirect \gls{IRI} estimation. Smartphone-based and embedded sensor systems rely on vehicle accelerometer data combined with signal processing and machine learning techniques \cite{XU2024}. Microcontroller-based \gls{IMU} platforms and sensor-fusion approaches have shown promising agreement with reference profilers \cite{LEI2018}\cite{ALATOOM2026}, and machine learning models such as random forests and neural networks have been used to map vehicle response data to \gls{IRI} values \cite{XU2024}.

Despite these advances, two important gaps remain. Most low-cost systems are validated only against reference \gls{IRI} measurements, without evaluating their relationship to passenger-perceived ride quality. Moreover, while several studies report high statistical agreement under controlled conditions, the interchangeability and consistency of different low-cost measurement approaches are still not fully understood.

This paper proposes a multi-modal, human-centric framework for low-cost roughness assessment based on three complementary sources: (i) subjective passenger evaluations, (ii) \gls{IRI} estimates from two independent smartphone applications, and (iii) raw in-vehicle \gls{IMU} measurements. Rather than relying exclusively on a high-precision ground-truth profiler, the study focuses on relative agreement, convergent validity, and perceptual relevance. The objective is to assess whether low-cost sensing can provide reliable intermediate evaluations to bridge the gap between official survey campaigns.

The main contributions of this work are: 1) A systematic comparison of two smartphone-based \gls{IRI} estimation methods in terms of agreement and consistency; 2) An analysis of the relationship between vehicle dynamic response (\gls{IMU} data) and app-derived \gls{IRI} estimates; 3) A quantitative investigation of the correlation between physical roughness proxies and passenger-perceived ride quality; 4) An assessment of the feasibility of using low-cost sensing for interim network-level roughness monitoring.

% By integrating infrastructure sensing and user perception, this work contributes to scalable, data-driven pavement monitoring strategies that complement periodic high-precision surveys and enhance the temporal resolution of road condition assessment.

The remainder of this paper is organised as follows: the next section describes related work in the field. Section~\ref{sec:experiment} describes the data collection process as well as the different data sources used in this work. Section~\ref{sec:methodology} presents an overview of methodology for the statistical analysis. Section~\ref{sec:results} presents and discusses the obtained results. Finally, Section~\ref{sec:conclusion} concludes the present work, outlining future research.
%%%%%%%%%%%%%%%%%%%%%%%%%%%%%%%%%%%%%%%%%%%%%%%%%%%%%%%%%%%%%%%%%%
%%%%%             RELATED WORK                            %%%%%%%%
%%%%%%%%%%%%%%%%%%%%%%%%%%%%%%%%%%%%%%%%%%%%%%%%%%%%%%%%%%%%%%%%%%
\section{Related Work}
\label{sec:relatedwork}
The \gls{IRI} is conventionally calculated from the longitudinal road profile using high-speed inertial profilers equipped with laser height sensors, accelerometers, and distance measurement instruments \cite{XU2024}. Although these systems provide highly accurate and repeatable measurements, their acquisition and operational costs limit their deployment frequency. As a result, \gls{IRI} surveys are typically conducted at multi-year intervals, creating temporal gaps in pavement condition knowledge. This limitation has motivated extensive research into low-cost and scalable alternatives capable of providing complementary roughness assessments.

A major research direction relies on vehicle response measurements, particularly vertical acceleration collected using smartphones or embedded \gls{IMU}. These response-based approaches infer roughness from dynamic vehicle behaviour rather than directly measuring pavement geometry. Numerous studies have demonstrated that acceleration-derived indicators can correlate with reference \gls{IRI} measurements under controlled conditions \cite{ALEADELAT2018}, \cite{ALSABAEEI2024}. However, because such systems measure the combined vehicle-road interaction, they are inherently sensitive to suspension characteristics, vehicle type, mounting position, and speed \cite{ALATOOM2026}, \cite{KATICHA2016}. Probe-vehicle investigations have shown that acceleration-derived roughness indicators strongly depend on sampling resolution and model assumptions, especially when approximating the standardised quarter-car formulation underlying \gls{IRI} \cite{KATICHA2016}.

To improve robustness, researchers have proposed both physics-based and data-driven strategies. Model-based approaches employ quarter-car or half-car dynamic models to estimate profile-related quantities or compute IRI-consistent indices from measured acceleration \cite{XU2024}, \cite{XUE2020}. Although these methods enhance interpretability, they require accurate knowledge of vehicle parameters and are sensitive to uncertainty in the modelling. In parallel, machine learning techniques have been widely adopted to map acceleration features, often including frequency-domain statistics, to reference \gls{IRI} values \cite{SANG2024}, \cite{ZHAO2024}. Smartphone-based systems using regression, ensemble models, and neural networks have reported promising accuracy when trained and validated within specific operating conditions \cite{MIRTABAR2022}. Nevertheless, sensor heterogeneity across smartphone models, orientation variability, and differences in mounting configuration remain key obstacles to generalisation \cite{ALSABAEEI2024}, \cite{AHMED2021}.

Beyond pure acceleration-based methods, alternative sensing modalities have been investigated. Vision-based approaches use cameras and convolutional neural networks to detect cracks and surface distress \cite{CHEN2022}. While effective for identifying localised defects, such systems estimate surface condition rather than longitudinal profile, requiring empirical correlation or multimodal fusion to approximate \gls{IRI}. Other low-cost concepts incorporate ultrasonic or infrared distance sensors to capture geometric information more directly \cite{BEHERA2021}, and sensor-fusion strategies combine \gls{IMU} measurements with additional sensing channels to reduce vehicle dependency \cite{ALATOOM2026}. These systems improve stability but introduce added hardware complexity and calibration requirements.

A complementary body of work focuses on network-level \gls{IRI} prediction using pavement age, traffic loading, structural properties, and environmental variables \cite{PEREZACEBO2023}. Although useful for pavement management forecasting, these models do not provide direct condition measurements and therefore complement rather than replace sensing-based approaches.

Despite substantial progress, most existing studies share two limitations. First, alternative methods are typically validated against a single high-precision reference profiler, implicitly assuming access to ground-truth \gls{IRI}. Second, relatively little attention has been given to the relationship between instrumented roughness metrics and passenger-perceived ride quality, even though IRI is historically rooted in vehicle response and ride comfort concepts. In real-world deployments where reference profilers are unavailable, the central question shifts from absolute accuracy to inter-method consistency and perceptual relevance.

The present study addresses these gaps by adopting a triangulation framework under no ground-truth conditions. Instead of treating one measurement source as authoritative, three complementary signals are analysed: subjective passenger ratings, \gls{IRI} estimates from two independent smartphone-based applications, and in-vehicle \gls{IMU}-derived dynamic features. This work extends existing response-based literature by evaluating inter-app agreement, examining the convergent validity between acceleration features and app-derived IRI, and assessing the association between physical roughness proxies and perceived comfort. The approach contributes toward perception-aware and multi-modal road condition assessment suitable for interim monitoring between specialised survey campaigns.

%%%%%%%%%%%%%%%%%%%%%%%%%%%%%%%%%%%%%%%%%%%%%%%%%%%%%%%%%%%%%%%%%
%%%%%             DDATA COLLECTION                 %%%%%%%%
%%%%%%%%%%%%%%%%%%%%%%%%%%%%%%%%%%%%%%%%%%%%%%%%%%%%%%%%%%%%%%%%%%
\section{Data Collection and Experimental Setup}
\label{sec:experiment}

\begin{figure}[tbp]
\centerline{\includegraphics[width=0.35\textwidth]{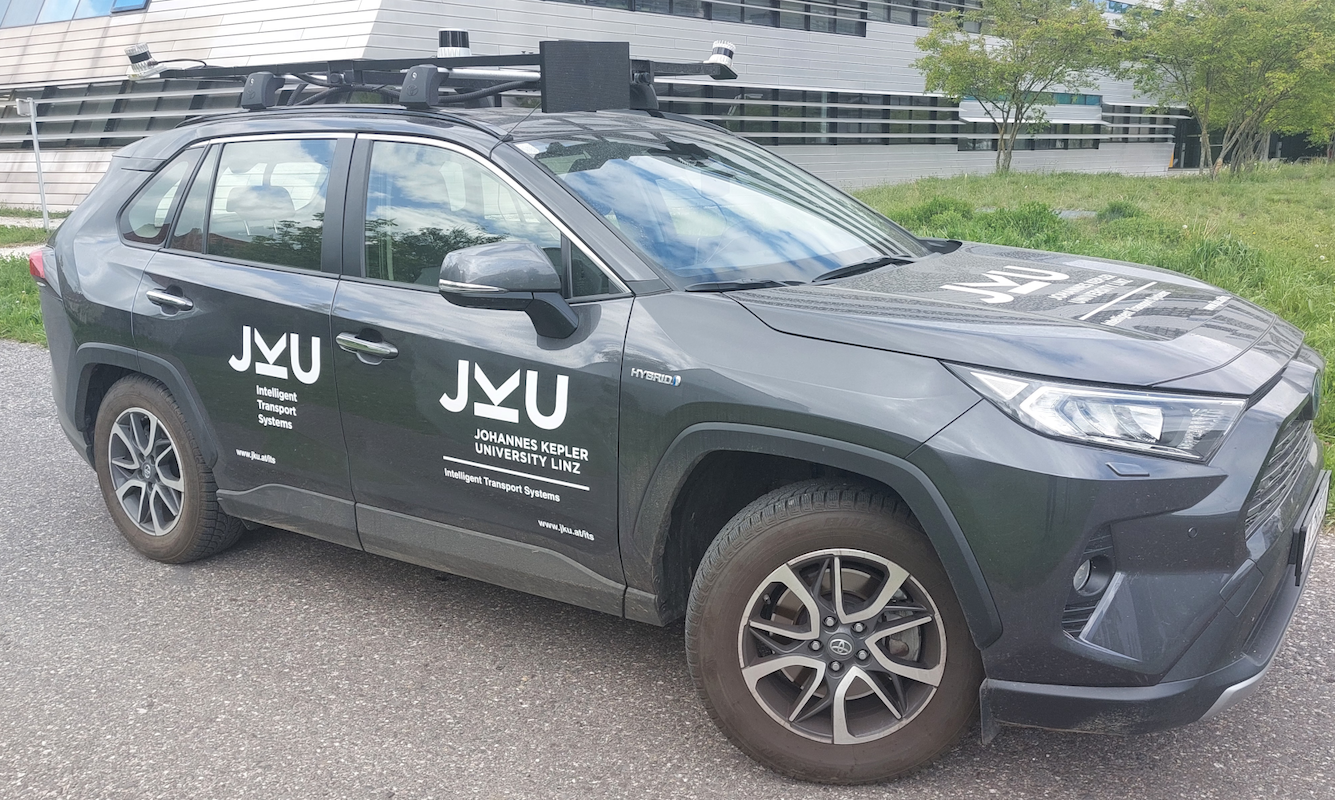}}
\caption{Vehicle used in the data collection \cite{its-vehicle}.}
\label{fig:vehicle}
\vspace{-0.5cm}
\end{figure}

The dataset was acquired using an instrumented passenger vehicle equipped with a multimodal sensing platform (as seen in Fig.~\ref{fig:vehicle}.) comprising a \gls{GNSS} receiver with \gls{IMU} (Xsens MTI-680G, RTK corrected when available), three LiDAR sensors (1xOS2 with 64 layers, 2xOS0 with 128 layers, 10 Hz), three monocular cameras (Basler, 10 fps), and CAN bus access logging signals such as linear speed per wheel, throttle and brake signal at 20Hz. All sensors were rigidly mounted to ensure mechanical stability and consistent extrinsic transformations between coordinate frames \cite{its-vehicle}. For the purpose of this study, only the \gls{IMU} and \gls{GNSS} data were used to extract features related to road roughness. Aiming at low-cost alternatives to dedicated survey vehicles, two additional data sources data sources were selected and used in parallel for data collection: a smartphone running two different apps to measure the \gls{IRI}; and the subjective perception of road quality assessed by the passengers inside the vehicle.

\begin{figure}[tbp]
\centerline{\includegraphics[width=0.30\textwidth]{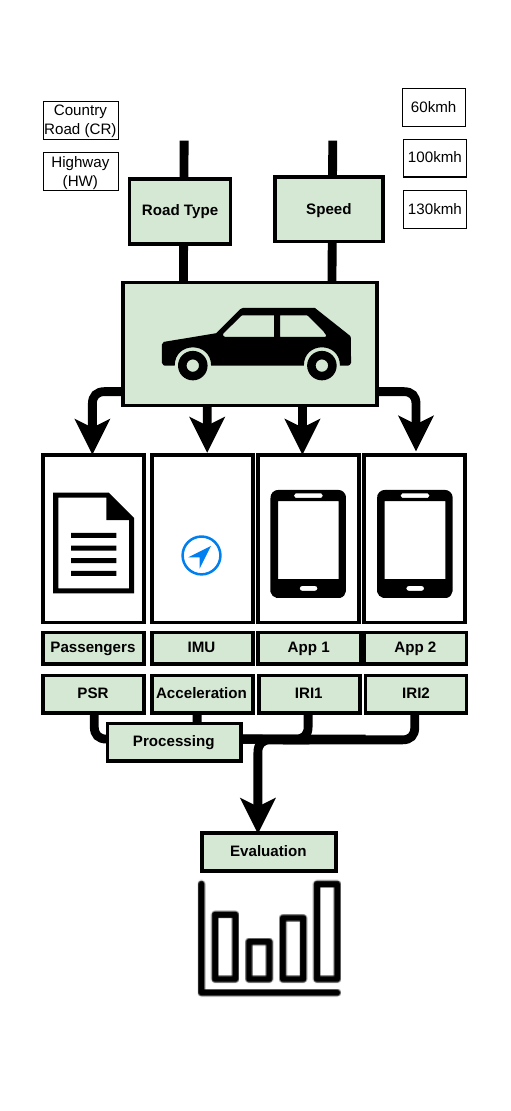}}
\vspace{-0.9cm}
\caption{General procedure followed during the data collection and the subsequent data processing.}
\label{fig:procedure}
\vspace{-0.2cm}
\end{figure}

The three data sources were collected at the same time along highways and interurban roads in a round trip between Linz (Austria) and Cluj (Romania), passing through Hungary. The trip covered approximately 1700km and lasted 30 hours, however, the recording was enabled $\approx25\%$ of the duration of the trip for a total of 7 hours recorded. The recordings were collected in segments of 12-15 minutes of continuous recording. Vehicle operation was conducted under normal traffic conditions and in compliance with local traffic regulations. The diversity in road types and countries ensures variability in pavement quality, structural characteristics, and environmental conditions. The general procedure is illustrated in Fig.~\ref{fig:procedure}. The full dataset is available at \cite{cemrr}. Details on each of the collected data sources are provided in the following subsections.

\subsection{In Vehicle GNSS/IMU }
\label{sec:dataset}

The \gls{IMU} provides tri-axial acceleration and angular velocity measurements at 200 Hz, while the \gls{GNSS} receiver logs position and velocity at 4 Hz. A fused pose is also calculated at 10Hz. The device was rigidly mounted to the vehicle's chassis to ensure consistent measurement of vertical dynamics associated with road roughness. The raw acceleration data were processed to compute the root mean square of the vertical acceleration ($Az\_rms$) for each road segment, which serves as a physical proxy for ride quality and pavement irregularities.

\subsection{Smartphone measurement of International Roughness Index}
% \begin{figure}[tbp]
% 	\centering
% 	\begin{subfigure}{0.237\textwidth}
% 		\includegraphics[width=\textwidth]{figures/app1_screenshot.jpeg}
% 		\caption{}
% 	\end{subfigure}
%     \begin{subfigure}{0.237\textwidth}
% 		\includegraphics[width=\textwidth]{figures/app2_screenshot.jpeg}
% 		\caption{}
% 	\end{subfigure}
% 	\caption{Screenshots of the two mobile applications used to measure road roughness. (a) Road IRI and (b) RoadLab Pro}
% 	\label{fig:apps}
%     \vspace{-0.5cm}
% \end{figure}

\label{sec:apps}
Road surface quality was independently evaluated using two mobile applications designed to estimate the \gls{IRI}:
% , as shown in Fig.~\ref{fig:apps}:
\begin{itemize}
	\item Application 1: Road IRI (Version 1.4.3) from Definitics Software Solution Pvt Ltd, India.
	\item Application 2: RoadLab Pro (Version 2.0.162) from SoftTeco, Europe.
\end{itemize}

The applications were executed on a Samsung Galaxy A56, mounted at the centre of the in-vehicle front panel. Each application computes IRI using the internal sensors of the smartphone (accelerometer and \gls{GNSS}) but the details of the models or algorithms used are not publicly available. \gls{IRI} measurements were segmented into road sections of approximately 100 m for further analysis.

\subsection{Present Serviceability Rating (PSR) }
To complement objective roughness metrics, subjective ride quality, defined as the perceived comfort of passengers during travel, was assessed using the \gls{PSR} \cite{XU2024}\cite{Gillespie1980Calibration}\cite{ALSABAEEI2024}. A panel of three raters provided ratings. Participants were briefed on the PSR methodology before starting the trip. \gls{PSR} was evaluated using a rating scale (1–5), where ``5'' stands for excellent road quality, and ``1'' means essentially impassable \cite{LEI2018}. Ratings were collected during driving, immediately after each segment, and correspond to road segments of approximately 8 km. The low number of raters was a practical choice due to the extension of the evaluated road and the availability of only one vehicle, but it also limits the generalisability of the subjective ratings. The PSR values were averaged across the three raters for each segment to obtain a single PSR score per segment for analysis.

\section{Analysis Methodology}
\label{sec:methodology}
\subsection{Data alignment and segmentation}
Since the datasets were collected independently, a first step was to align the data sources in time and space. The \gls{GNSS} data from the vehicle and the smartphone were used to synchronise the datasets into consistent road sections. Two smartphone applications provided roughness estimates at 100 m resolution, denoted as $IRI_1$ and $IRI_2$. Passenger feedback was collected as a \gls{PSR} on an numerical scale (1–5) every 5 minutes. To ensure comparability across sources, all streams were mapped to a common set of road segments. Let $s$ index analysis segments. Each segment $s$ is defined by a spatial window (300 m) and associated metadata: road type $r(s)$ (highway, main road) and mean speed $v(s)$. For each segment, the following metrics were computed:%we computed:

\begin{itemize}
	\item $IRI_1(s)$ and $IRI_2(s)$: Average \gls{IRI} values from the two apps for that segment, obtained by averaging the app outputs that fall within the segment's spatial window. If no app output is available for a segment, the value is marked as missing.
	\item $PSR(s)$: PSR mapped to the segment based on timestamp overlap with the 5-minute rating window. When multiple segments fall within one PSR window, the same PSR value is assigned to all segments in that window. The PSR value is the average of the ratings provided by the three participants.
	\item $Az\_rms(s)$: Root mean square of the vertical acceleration measured by the \gls{IMU} in the vehicle.
\end{itemize}

\subsection{Correlation Analysis}

Linear association between the two variables was quantified using the Pearson correlation coefficient ($r$). Let $x_s$ and $y_s$ denote the values of the two variables for segment $s$, for $n$ matched segments. Pearson's correlation coefficient is defined as:

\begin{equation}
r_{xy} =\frac{\sum_{s=1}^{n} (x_s - \bar{x})(y_s - \bar{y})}{
\sqrt{\sum_{s=1}^{n} (x_s - \bar{x})^2}
\sqrt{\sum_{s=1}^{n} (y_s - \bar{y})^2}},
\end{equation}

where

\begin{equation}
\bar{x} = \frac{1}{n} \sum_{s=1}^{n} x_s,\qquad\bar{y} = \frac{1}{n} \sum_{s=1}^{n} y_s.
\end{equation}

 The null hypothesis that the distributions underlying both variables are uncorrelated and normally distributed was also tested and the $p$-values were reported. Six different $x_s/y_s$ pairs were considered in the correlation study: $\frac{\mathrm{IRI}_1}{\mathrm{IRI}_2}$, $\frac{\mathrm{IRI}_1}{\mathrm{Az\_rms}}$,  $\frac{\mathrm{IRI}_2}{\mathrm{Az\_rms}}$, $\frac{\mathrm{IRI}_1}{\mathrm{PSR}}$, $\frac{\mathrm{IRI}_2}{\mathrm{PSR}}$, and $\frac{\mathrm{Az\_rms}}{\mathrm{PSR}}$.

\subsection{Agreement Analysis}

Agreement analysis assesses the degree of concordance between two (in this case) sets of measurements to determine whether they can be used interchangeably, which can no be determine solely by correlation studies \cite{RANGANATHAN2017}. Due to the lack of ground truth values for IRI (gold standard), the agreement between $\mathrm{IRI}_1$ and $\mathrm{IRI}_2$ was assessed using the \gls{ICC} and Bland--Altman analysis \cite{RANGANATHAN2017}\cite{TAFFE2021}.

\subsubsection{Bland--Altman Analysis}

For each segment $s$, the pairwise mean and difference were computed as:

\begin{equation}
m_s =\frac{\mathrm{IRI}_1(s) + \mathrm{IRI}_2(s)}{2},\qquad d_s =\mathrm{IRI}_1(s) -\mathrm{IRI}_2(s).
\end{equation}

The estimated bias (mean difference) is:

\begin{equation}
\bar{d} =\frac{1}{n} \sum_{s=1}^{n} d_s,
\end{equation}

and the standard deviation of the differences is:

\begin{equation}
s_d =\sqrt{\frac{1}{n-1}\sum_{s=1}^{n}(d_s - \bar{d})^2
}.
\end{equation}

The limits of agreement (LoA) were calculated as:

\begin{equation}
\mathrm{LoA}_{\pm} =\bar{d} \pm 1.96\, s_d.
\end{equation}

Bland--Altman plots display $d_s$ versus $m_s$, together with horizontal lines corresponding to $\bar{d}$ and $\mathrm{LoA}_{\pm}$.

To assess proportional bias, a linear regression model was optionally fitted:

\begin{equation}
d_s = \beta_0 + \beta_1 m_s + \varepsilon_s,
\end{equation}

where a statistically significant $\beta_1$ indicates magnitude-dependent disagreement.

\subsubsection{Intraclass Correlation Coefficient}

A two-way mixed-effects model with absolute agreement (ICC(A,1)) was adopted. Let $Y_{si}$ denote the IRI value for segment $s$ measured by application $i \in \{1,2\}$. With $n$ segments and $k=2$ raters, the ICC(A,1) is defined as:

\begin{equation}
\mathrm{ICC(A,1)} =\frac{MS_R - MS_E}{MS_R + (k-1)MS_E + \frac{k}{n}(MS_C - MS_E)},
\end{equation}

where:

\begin{itemize}
    \item $MS_R$ is the mean square for rows (segments),
    \item $MS_C$ is the mean square for columns (applications),
    \item $MS_E$ is the residual mean square.
\end{itemize}

ICC values were computed for the complete dataset and within each road-type and speed-based subgroup.

\subsection{Statistical Considerations}

%All analyses were conducted using matched segment pairs only. Statistical significance was evaluated at a significance level of $\alpha = 0.05$. Correlation coefficients, ICC values, and Bland--Altman parameters were reported both globally and within each stratified subgroup. As in:
All analyses were conducted using matched segment pairs only. Statistical significance was evaluated at a significance level of $\alpha = 0.05$. The dataset was analysed both as a whole and after division into stratified subgroups. Correlation coefficients, ICC values, and Bland--Altman parameters were reported for the entire dataset as well as for each subgroup. A summary of the composition of each subgroup is shown in Table~\ref{table:data}, where the stratification was defined as follows:
\begin{itemize}
    \item on the full dataset: (All),
    \item stratified by country: Austria (AT), Hungary (HU), and Romania (RO),
    \item stratified by road type: Highway(HW), and Main Road (MR),
    \item stratified by road type and speed intervals: Highways were further categorised into HW60 (speed $<60$ km/h), HW100 (speed between 60 and 100 km/h), and HW130 (speed $> 100$ km/h). Main roads were categorised into MR60 (speed $< 60$ km/h) and MR100 (speed $\geq 60$ km/h).
\end{itemize}
%A summary of the dataset stratified by average values per segment is shown in Table~\ref{table:data}.

\begin{table}[t]
\caption{Dataset summary. Distance covered, percentage of highway (HW) and main road (MR) segments, mean \gls{IRI} from both applications, and mean \gls{PSR} rating per condition.}%\textcolor{red}{update HW and CR percentages per country}}
\label{table:data}
\begin{center}
\resizebox{\columnwidth}{!}{
\begin{tabular}{cccccccc}
\toprule
\multirow{2}{*}{Dataset} & $v$ & Distance & HW & CR & $IRI_1$ & $IRI_2$ & \multirow{2}{*}{PSR} \\
        & [km/h]        & [km] & [\%] & [\%]  & [m/km] & [m/km] & \\
\midrule
All	    & 87 	& 381 	& 71\% 	& 29\% 	& 1.15 & 4.26 & 3.7 \\	
AT      & 101 	& 58 	& 100\% & 0\% 	& 1.17 & 4.69 & 4.2 \\		
HU      & 93 	& 226 	& 57\% 	& 43\% 	& 1.11 & 3.82 & 3.8 \\		
RO      & 65 	& 97  	& 17\%  & 83\% 	& 1.24 & 5.05 & 3.5 \\		
HW      & 101 	& 205 	& 100\% & 0\% 	& 1.10 & 3.52 & 4.2 \\	
HW60    & 48 	& 5 	& 100\% & 0\% 	& 1.22 & 4.76 & - \\	
HW100   & 86 	& 78 	& 100\% & 0\% 	& 1.13 & 3.71 & - \\	
HW130   & 112 	& 122 	& 100\% & 0\% 	& 1.07 & 3.35 & - \\	
MR      & 71 	& 176 	& 0\% 	& 100\% & 1.22 & 5.13 & 3.2 \\	
MR60    & 51 	& 58  	& 0\% 	& 100\% & 1.33 & 5.16 & -\\	
MR100   & 80 	& 118  	& 0\% 	& 100\% & 1.16 & 5.12 & - \\	
\bottomrule
\end{tabular}
}
\end{center}
\vspace{-0.5cm}
\end{table}

%%%%%%%%%%%%%%%%%%%%%%%%%%%%%%%%%%%%%%%%%%%%%%%%%%%%%%%%%%%%%%%%%%
%%%%%                  RESULTS                            %%%%%%%%
%%%%%%%%%%%%%%%%%%%%%%%%%%%%%%%%%%%%%%%%%%%%%%%%%%%%%%%%%%%%%%%%%%
\section{Results}
\label{sec:results}
The general results of the correlation and agreement analyses between $\mathrm{IRI}_1$, $\mathrm{IRI}_2$, $\mathrm{Az}_{rms}$, and $\mathrm{PSR}$ across different conditions are summarised in Table~\ref{table:results}. For each pairwise comparison, Pearson correlation coefficient ($r$), and $p$-values were reported. The \gls{ICC}, and their corresponding $p$-values and 95\% confidence intervals (CI95) are also reported. The results are presented both for the entire dataset (All) and for specific subgroups based on country, road type, amd speed.
\begin{table*}[t]
\caption{Summary of correlation and agreement analyses between $\mathrm{IRI}_1$, $\mathrm{IRI}_2$, $\mathrm{Az}_{rms}$, and $\mathrm{PSR}$ across different conditions}
\label{table:results}
\begin{center}
\resizebox{\textwidth}{!}{
\begin{tabular}{c|cccccc|cc|cc|cc|cc|cc}	
\toprule
\multirow{2}{*}{Dataset} & \multicolumn{6}{c}{$\frac{IRI_1}{IRI_2}$} & \multicolumn{2}{c}{$\frac{IRI_1}{Az_{rms}}$} &  \multicolumn{2}{c}{$\frac{IRI_2}{Az_{rms}}$} & \multicolumn{2}{c}{$\frac{IRI_1}{PSR}$} &  \multicolumn{2}{c}{$\frac{IRI_2}{PSR}$} & \multicolumn{2}{c}{$\frac{Az_{rms}}{PSR}$} \\[1.8ex]
%\cmidrule{4-9}
 & $r$  & $p$ & $\beta_1$ & $ICC$  & $p$ & CI95 & $r$ & $p$ & $r$ & $p$& $r$ & $p$ & $r$ & $p$ & $r$ & $p$\\
\midrule
all 	&0.72	&$<0.001$	& 0.89	& 0.22	&$<0.001$&  [0.15, 0.28]	&-0.12 	&$<0.001$&-0.11 &0.001		&0.09 &0.145		&0.11 	&0.010 		&-0.49 	&0.030\\
AT  	&0.72	&$<0.001$	& 0.88	& 0.24	& 0.005 &  [0.06, 0.4]		&0.17 	&0.017	 &0.21 	&0.002		&0.12 	&0.020		&0.10 	&0.007 		&-0.46	&0.017\\
HU 		&0.75	&$<0.001$	& 0.90	& 0.21	&$<0.001$&  [0.12, 0.29]	&-0.14 	&0.003   &-0.12 &0.009		&-0.12 &0.003		&-0.15 	&0.010 		&-0.50 	&0.025\\
RO 		&0.69	&$<0.001$	& 0.85	& 0.41	&$<0.001$&  [0.28, 0.52]	&-0.27 	&$<0.001$&-0.34 &$<0.001$	&0.40 &0.086		&0.43 	&$<0.001$ 	&-0.45 	&0.013\\
HW 		&0.65	&$<0.001$	& 0.87	& 0.17	&$<0.001$&  [0.07, 0.26]	&0.07 	&0.114	 &0.10 	&0.021		&0.20 	&0.152		&0.17 	&0.040 		&-0.39	&0.041\\
HW60 	&0.16	&0.672		& 0.96	& 0.05	& 0.443 &  [-0.6, 0.66]		&-0.55 	&0.159   &-0.25 &0.555		&-0.42 &0.112		&-0.50 	&0.239 		& - 	& - \\
HW100 	&0.69	&$<0.001$	& 0.87	& 0.21	& 0.004 &  [0.06, 0.36]		&-0.18 	&0.024   &-0.18 &0.025		&0.19 &0.019		&0.10 	&0.027 		& -	& - \\
HW130 	&0.66	&$<0.001$	& 0.88	& 0.15	& 0.009 &  [0.03, 0.27]		&0.27 	&$<0.001$&0.36 	&$<0.001$	&0.26 	&$<0.001$	&0.23 	&$<0.001$ 	& -	& - \\
MR 		&0.78	&$<0.001$	& 0.90	& 0.30	&$<0.001$&  [0.2, 0.39]		&-0.11 	&0.059   &-0.11 &0.051		&0.15 	&0.082		&0.09 	&0.060 		&-0.44 	&0.023\\
MR60 	&0.79	&$<0.001$	& 0.91	& 0.41	&$<0.001$&  [0.25, 0.55]	&0.06 	&0.529	 &0.12 	&0.217		&-0.07 	&0.458		&0.02 	&0.174 		& -	& - \\
MR100 	&0.84	&$<0.001$	& 0.88	& 0.27	&$<0.001$&  [0.15, 0.38]	&0.04 	&0.562	 &0.07 	&0.314		&0.10 	&0.372		&0.06 	&0.344 		& -	& - \\

% \bottomrule
\end{tabular}
}
\end{center}
\vspace{-0.5cm}
\end{table*}

\begin{figure}[tbp]
\centerline{\includegraphics[width=0.48\textwidth]{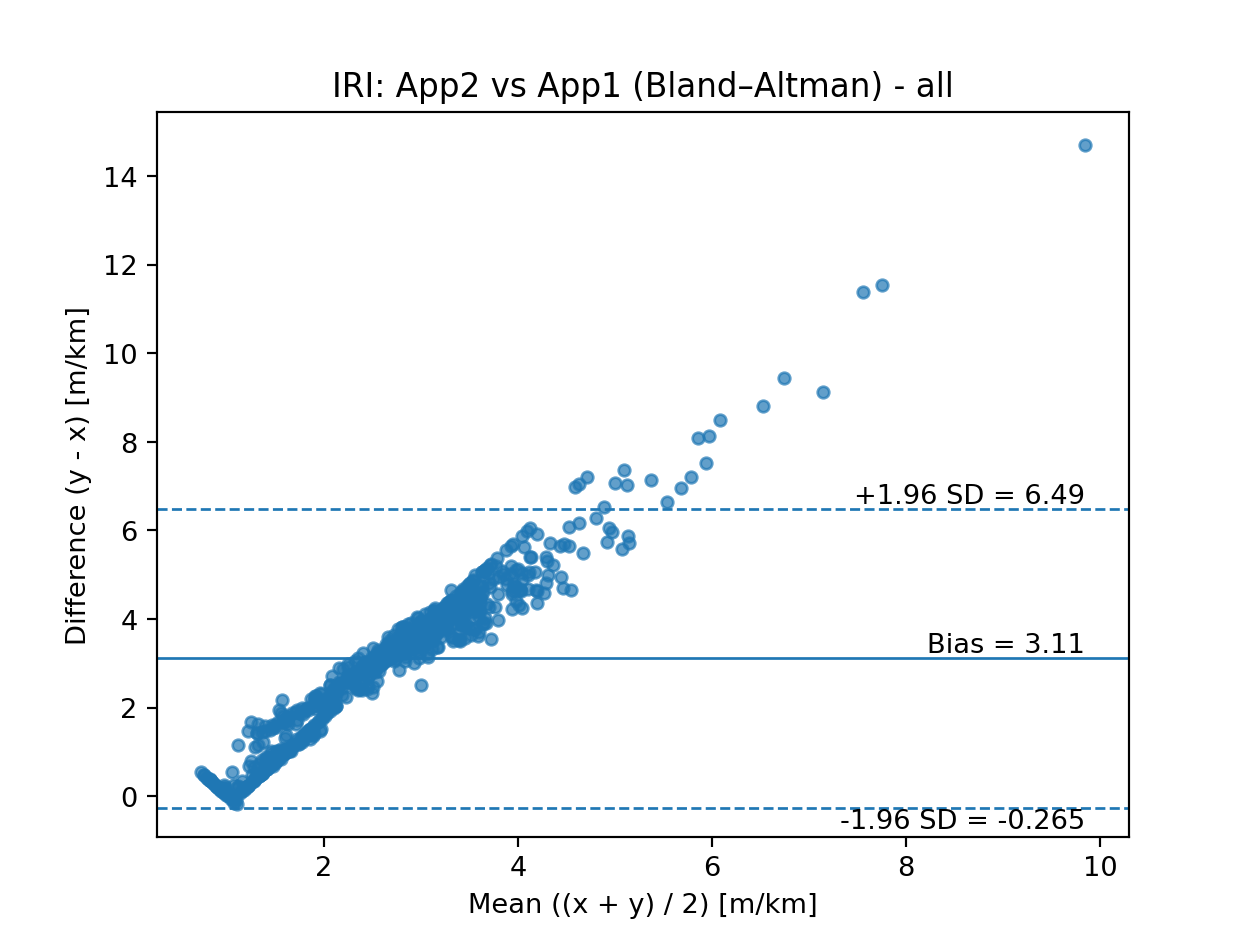}}
% \vspace{-0.9cm}
\caption{Band--Altman plot for the comparison between $\mathrm{IRI}_1$ and $\mathrm{IRI}_2$ across the whole dataset. The solid line represents the mean difference ($\bar{d}$), while the dashed lines indicate the limits of agreement (LoA). Each point corresponds to a segment, with its position determined by the mean IRI value from both applications (x-axis) and their difference (y-axis). The values exhibit a strong proportional bias ( $\beta_1$ = 0.89).}
\label{fig:band-altman}
\vspace{-0.2cm}
\end{figure}

The condition HW60 was excluded from analyses due to the low number of segments resulted in non-significant correlations and unreliable agreement metrics.

Regarding the comparison between $IRI_1$ and $IRI_2$, high correlation ($r > 0.6$) is consistent across all the conditions. However, the Band--Altman analysis shows a strong proportional bias, with a significant positive slope across all the studied conditions. The slope values ($\beta_1$) are reported in Table~\ref{table:results}. The rest of the parameters of the Band--Altman plot ($\bar{d}$, $s_d$, $\mathrm{LoA}$) were not reported because they are unusable for such case. As an exemplary case, the Bland--Altman plot for the entire dataset is shown in Fig.~\ref{fig:band-altman}, clearly exhibiting a positive slope $\beta_1 = 0.89$, indicating proportional bias where disagreement increases with higher \gls{IRI} values. Low \gls{ICC} values (0.15-0.41) indicate poor to moderate agreement, suggesting that the two applications cannot be used interchangeably without calibration.

$IRI_1$ and $IRI_2$ show weak and inconsistent correlations with $Az_{rms}$ across conditions, with some significant negative and positive correlations (e.g., in RO, HW130) and others showing no significant association. This suggests that the IRI estimations may not consistently reflect the physical roughness as measured by vertical acceleration, potentially due to differences in sensor processing , sensitivity to specific road features, or mounting issues.

On the comparison of $IRI_1$ and $IRI_2$ and \gls{PSR}, again the correlations are weak and inconsistent across conditions, with some significant positive and negative correlations. This suggests that the IRI estimations may not consistently reflect the perceived ride quality, potentially due to differences in how each app processes sensor data, or because PSR is influenced by factors beyond roughness (e.g., noise, traffic conditions).

On the contrary, $Az_{rms}$ shows a consistent and significant negative correlation with PSR across all conditions, confirming that higher vertical acceleration (indicating rougher roads) is associated with lower perceived ride quality. Unfortunately, due to low number of ratings, it was not possible to segment the PSR data by speed, which limits the analysis of how these factors may influence the relationship between physical roughness and perceived comfort.

%%%%%%%%%%%%%%%%%%%%%%%%%%%%%%%%%%%%%%%%%%%%%%%%%%%%%%%%%%%%%%%%%%
%%%%%                  CONCLUSION                          %%%%%%%
%%%%%%%%%%%%%%%%%%%%%%%%%%%%%%%%%%%%%%%%%%%%%%%%%%%%%%%%%%%%%%%%%%
\section{Conclusion and Future Work}
\label{sec:conclusion}
%\section{Conclusion}

This study proposed and evaluated a multi-modal framework for road roughness assessment combining smartphone-based IRI estimation, in-vehicle dynamic measurements, and passenger Present Serviceability Ratings (PSR). Rather than relying on a single measurement modality, the analysis examined consistency between applications, physical dynamic response, and perceived ride quality when ground truth is not available. Although the two smartphone applications exhibited strong linear correlation across all analysed conditions, agreement analysis revealed poor to moderate absolute reliability and a pronounced proportional bias. These findings indicate that, despite following similar trends, the applications cannot be considered interchangeable and may produce systematically divergent roughness values at higher IRI levels. More importantly, the relationship between app-derived IRI and PSR was weak and inconsistent across countries, road types, and speed regimes. In contrast, vertical acceleration demonstrated a consistent and statistically significant negative correlation with PSR across all analysed conditions. This result confirms that increased vehicle dynamic excitation is directly associated with reduced perceived ride quality.

The strong and stable association between vertical acceleration and PSR suggests that passenger perception captures physically meaningful roughness effects and can serve as a valid and sensitive indicator of road surface condition. Future work will focus on multi-vehicle validation and the development of predictive models integrating infrastructure, dynamic response, and human perception.

%%%%%%%%%%%%%%%%%%%%%%%%%%%%%%%%%%%%%%%%%%%%%%%%%%%%%%%%%%%%%%%%%%
%\vfill
\section*{ACKNOWLEDGMENT}
The work has been conducted as a part of MITHOS project (No.101202959) funded by the European Union. Views and opinions expressed are however those of the author(s) only and do not necessarily reflect those of the European Union or CINEA. Neither the European Union nor the granting authority can be held responsible for them.

%%%%%%%%%%%%%%%%%%%%%%%%%%%%%%%%%%%%%%%%%%%%%%%%%%%%%%%%%%%%%%%%%%%%
%\addtolength{\textheight}{-12cm}
%\vspace{10mm}
\bibliographystyle{IEEEtran}
\bibliography{paper}

\end{document}